\documentstyle[epsf]{article}
\begin{document}

\begin{center}
{\Large \bf Jet observables and energy-momentum tensor} \\

\vspace{8mm}

P.S.Cherzor and N.A.Sveshnikov\footnote[2]{Deceased.}\\ 
\vspace{4mm}
Department of Physics, Moscow State University, Moscow, 119899\\ 
\end{center}

\vspace{8mm}

{ }

{\it Abstract.} 
We clarify and extend the theorem of Sveshnikov and
Tkachov \cite{svesh-95}, \cite{svesh-tka-96}, which gives an explicit 
connection between jet 
observables and the energy-momentum tensor. We check the relation between 
jet
observables and the energy-momentum tensor for non-scalar (spinor and 
vector)
fields, give a correct treatment of the light-cone 
singularity
for massless particles, and extend the theorem of \cite{svesh-95}, 
\cite{svesh-tka-96} to the 
massless
case. We also discuss the issue of gauge invariance.

\vspace{8mm}

\section{Introduction}

Modern QCD increasingly emphasizes precision measurements
and perturbative calculations of higher order corrections 
(cf. measurements of $\alpha_s$
and other parameters of the Standard Model). 

It was argued in \cite{tka-95} that in the context of precision 
measurements a central role is played by
a special class of observables --- the so-called C-correlators --- 
that contain all information about 
multijet structure and possess optimal stability properties with respect to 
experimental errors.
As was shown in \cite{svesh-95}, \cite{svesh-tka-96}, the C-correlators 
possess another property that makes 
them
extremely attractive from theoretical point of view. Namely, the theorem of 
Sveshnikov and Tkachov \cite{svesh-95}, \cite{svesh-tka-96} 
expresses C-correlators (and, consequently, a vast class of other jet 
observables \cite{tka-95}) in terms of the 
energy-momentum tensor 
in such a manner that no information about hadron bound states is used. 

In the arguments of \cite{svesh-95}, \cite{svesh-tka-96} there 
are some gaps. The purpose 
of this work is to
clarify them:

(i) The theorem was proved in \cite{svesh-95}, \cite{svesh-tka-96} only for 
scalar fields. So one needs to 
check
it for non-scalar spinor and vector fields. 

(ii) The theorem was accurately proved only for massive particles. The 
massless case exhibits some 
subtleties due to the light-cone singularity that have to be clarified. 

(iii) The issue of gauge invariance in the 
case
of gluons has to be clarified.

\section{Setup}

The 
C-correlators have the following form: 
\begin{equation}
\label{c0}\left\langle \sum\limits_{i_1}\ldots \
\sum\limits_{i_N}E_{i_1}\ldots E_{i_N\ }\ f_N\ (\ \widehat{{\bf p}}%
_{i_1},\ldots \widehat{{\bf p}}_{i_N}\ )\right\rangle _P. 
\end{equation}
Here $i_k$ are indexes that run over all particles produced in an event, N 
is the order of correlator $(N=1, 2,\ldots)$. Note that the 
definition is 
entirely in terms
of observable quantities --- particles' energies $(E_{i})$ and angles $( 
\widehat{{\bf p}}_{i})$.

One can rewrite it in the Fock space formalism 
as follows:
\begin{equation}
\label{c1}\int d{\bf n}_1\ldots \int d{\bf n}_N\ \left\langle in\left|
\varepsilon \ ({\bf n}_1)\ldots \varepsilon \ ({\bf n}_N)\right|
in\right\rangle \times f_N\ ({\bf n}_1,\ldots {\bf n}_N), 
\end{equation}
where $\varepsilon \ ({\bf n}_i)$ is an operator-valued distribution on the
unit sphere: 
\begin{equation}
\label{form008}
\varepsilon \ ({\bf n})\equiv \sum \int \frac{d{\bf 
p}}{2p_0}%
\,\,|{\bf p}|\ a^{+}({\bf p})\ a^{-}({\bf p})\,\delta (\ \widehat{{\bf 
p}},%
{\bf n}). 
\end{equation}
In \cite{svesh-95}, \cite{svesh-tka-96} the following relation 
between $\varepsilon \ ({\bf 
n})$ and
the energy-momentum tensor was found:

{\bf Theorem} (Sveshnikov and Tkachov, 1995): 
\begin{equation}
\label{form0}\varepsilon \ ({\bf n})=\lim \limits_{t\rightarrow \infty
}\;t^3\int\limits_0^1\rho ^2d\rho \ n_iT_{i0}(\rho {\bf n}t,t)\, ,
\end{equation} 
where $ T_{i0}$ is the energy-momentum tensor 
(the weak limit is implied here). 

In the context of high energy QCD we can regard gluon, quark and ghost
fields as massless. This corresponds to high energy limit $({\bf 
p,n)=}\left| 
{\bf p}\right| \simeq p_0$.

In order to compare the massive and massless cases, we will reverse 
the procedure of \cite{svesh-95}, \cite{svesh-tka-96}
and from the
field representation of $T_{i0}$ obtain operator representation for both
cases.

\section{Massive case}

\subsection{Scalar field}

We already mentioned the result for scalar massive fields. Let us start 
from the well-known 
massive scalar 
Lagrangian and
represent the fields in operator form: 
\begin{equation}
\label{d34}\varphi \;(x)=\frac 1{(2\pi )^{3/2}}\int \frac{d{\bf p}}{2p_0}%
(e^{-ipx\ }a^{-}({\bf p})+H.C.)\,. 
\end{equation}
Following \cite{svesh-95}, \cite{svesh-tka-96} one gets the following 
expression (in 
non-commutative
case):
\begin{equation}
\label{c88}\varepsilon \ ({\bf n})=\int\limits_0^\infty \frac{p^3dp}{4p_0}%
\left( a^{+}a^{-}(p{\bf n})+a^{-}a^{+}(p{\bf n})\right) . 
\end{equation}

\subsection{Spinor field}

Now we turn to massive spinor fields for which we will 
obtain a similar result. 
We start with free asymptotic fields. Write 
\begin{equation}
\label{form1}\psi \;(x)=\frac 1{(2\pi )^{3/2}}\int \frac{d{\bf p}}{2p_0}%
(e^{-ipx\ }u_s({\bf p})\ b_s({\bf p})\ +e^{ipx\ }v_s({\bf p})\ d_s^{+}(%
{\bf p})\ ) .
\end{equation}
The expression for the energy-momentum tensor for free fields is well-known 
(see e.g. \cite{bog-shir}). So let 
us substitute (\ref{form1}) into (\ref{form0}) and use the stationary
phase method. Only case of opposite signs in stationary phase equations 
contributes to 
$\varepsilon \ ({\bf n})$, 
because the same-sign contribution gives
asymptoticaly $\varepsilon \ ({\bf n})=0$. By straightforward substitution 
(\ref{form0}) 
one obtains the following formula:
$$
\lim \limits_{t\rightarrow \infty }\,\,t^3\int\limits_0^1\rho ^2d\rho \ 
{n_i} \frac
i2\left[ \overline{\psi }\gamma _i\partial _0\psi -\partial 
_0\overline{\psi 
}\gamma _i\psi \right] =\lim \limits_{t\rightarrow \infty }\frac 1{(2\pi
)^3}\int\limits_0^1\rho ^2d\rho \ {n_i} \int \frac{d{\bf p}}{2p_0}\int 
\frac{d{\bf 
q}%
}{2q_0} 
$$
\begin{equation}
\label{c9}\times \exp \{\pm \frac{i\sqrt{1-\rho ^2}}{2m}[({\bf p})^2-\rho^2 
(%
{\bf n},{\bf p})^2]\mp \frac{i\sqrt{1-\rho ^2}}{2m}[({\bf q})^2-\rho^2 
({\bf 
n}%
,{\bf q})^2]\}\cdot F, 
\end{equation}
$$
F=\frac{q_0}2[\overline{u}_s({\bf p}^{\prime })b_s^{+}({\bf p}%
^{\prime })\gamma _iu_s({\bf q}^{\prime })b_s({\bf q}^{\prime })-%
\overline{v}_s({\bf p}^{\prime })d_s({\bf p}^{\prime })\gamma _iv_s(%
{\bf q}^{\prime })d_s^{+}({\bf q}^{\prime })] 
$$
\begin{equation}
\label{c10}+\frac{p_0}2[\overline{u}_s({\bf p}^{\prime })b_s^{+}({\bf 
p}%
^{\prime })\gamma _iu_s({\bf q}^{\prime })b_s({\bf q}^{\prime })-%
\overline{v}_s({\bf p}^{\prime })d_s({\bf p}^{\prime })\gamma _iv_s(%
{\bf q}^{\prime })d_s^{+}({\bf q}^{\prime })]{,} 
\end{equation}
where 
\begin{equation}
\label{form2''}{\bf p}^{\prime }=\frac{m\rho }{\sqrt{1-\rho ^2}}{\bf 
n}+%
\frac{{\bf p}}{\sqrt{t}},\,\,\quad {\bf q}^{\prime }=\frac{m\rho 
}{\sqrt{%
1-\rho ^2}}{\bf n}+\frac{{\bf q}}{\sqrt{t}}.
\end{equation}
The difference from the scalar case is only in the operator part of the 
expression.
After rewriting $\varepsilon \;({\bf n})$ in terms of 
\begin{equation}
\label{c110}p\equiv \left| {\bf p}\right| =\frac{m\rho }{\sqrt{1-\rho ^2}}
\end{equation}
and using for the operator part of expression (\ref{c9}) the following 
formulae:
\begin{equation}
\label{c11}{\ }\overline{u}_s\gamma _\mu u_s=2p_\mu ,\quad \quad 
\overline{v}%
_s\gamma _\mu v_s=2p_\mu ,\quad \quad [d_s,d_s^{+}]_{+}=0 
\end{equation}
one obtains the final result: 
\begin{equation}
\label{x2}\int\limits_0^\infty \frac{p^3dp}{2p_0}(\;b_s^{+}b_s(p{\bf n)}%
+d_s^{+}d_s(p{\bf n)})\,,
\end{equation}
which agrees with the scalar case.

\subsection{Vector field}

The massive vector field is treated similarly. One starts with the
standard representation:
\begin{equation}
\label{C12}A_\mu (x)=\frac 1{(2\pi )^{3/2}}\int \frac{d{\bf p}}{2p_0}%
(e^{-ipx}A_\mu ^{-}({\bf p})+H.C.). 
\end{equation}
The vector field Lagrangian in general covariant arbitrary gauge is as 
follows: 
\begin{equation}
\label{r76}L=-\frac 12\partial ^\mu A^\nu \partial _\mu A_\nu +\frac{m^2}%
2A^\mu A_\mu +\frac \alpha 2\partial ^\mu A^\nu \partial _\nu A_\mu . 
\end{equation}
As a consequence of the Noether theorem it is always possible to add 
a
4-divergence to $T_{\mu \nu }$, which allows us to 
make a convenient choice of the Lagrangian. 
We
starts with the St\"uckelberg Lagrangian ($\alpha =1$ in (\ref{r76})) and
use the fact that the theory in this case is gauge invariant. The Lorentz
condition is applied (for $\alpha =1$ it 
appears automatically from the
Lagrangian) in the operator form (which arises from fields structure). 
In 
this way one
can avoid non-physical states. In this way there is no contradiction: 
one
retains Lorentz invariance and local commutativity. 
But the components $A_\mu^{\pm }({\bf p})$ 
are no longer independent.

The negative-sign contributions to the $T_{00}$ can be eliminated by
representing operators $A_\mu ^{\pm }({\bf p})$ in (\ref{C12}) 
in a local
frame of reference in momentum space (cf.\ e.g.\ \cite{bog-shir}). 
To obtain $T_{0i}$ one can choose any gauge 
for example $\alpha =0$
for calculational simplicity. 
The result is gauge independent and has the
form of (\ref{form008}), as expected. 
In the final formula only physical components survive: 
\begin{equation}
\label{x1}\int\limits_0^\infty \frac{p^3dp}{4p_0}(a_i^{+}a_i^{-}(p{\bf n)}%
+a_i^{-}a_i^{+}(p{\bf n)}){.} 
\end{equation}

\section{Massless case}

The result (\ref{form0}) has to be interpreted carefully in the massless 
case. From the
accurate derivation (given below) it follows that the integration from $0$
to $1$ should be spread (formally integration with $\theta $-function over 
$\rho $) over region A (see Fig.1 on the next page).


\begin{figure}[thb]
\centerline{\epsfbox{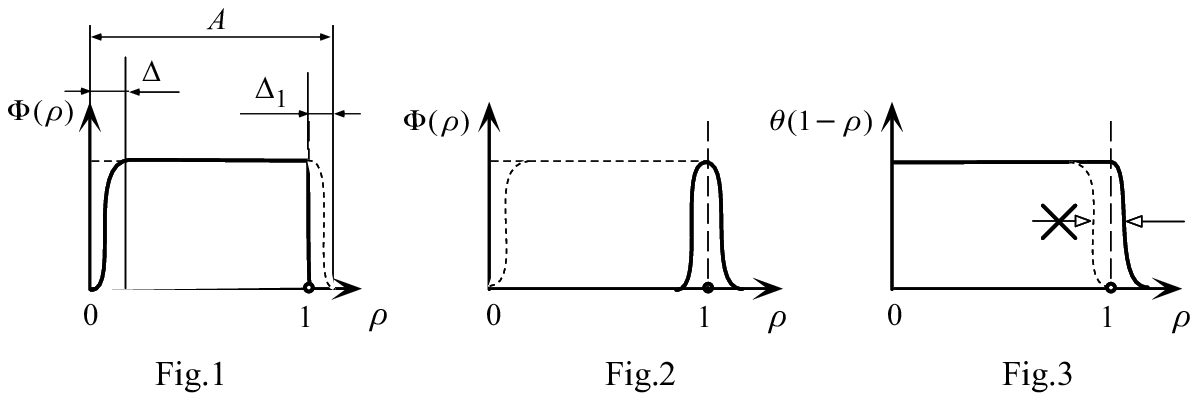}}
\end{figure}

At infinitesimal $\rho $ we can neglect contribution of particles in 
the $\Delta-$ region, 
because of limited experimental sensitivity and hence such slow
particles could not be registered by detectors. In the massive case, the
region $\Delta _1$ simply does not contribute as follows from the 
formula for
stationary phase manifold: in the denominator square root of negative
value does not have any physical meaning. But presence of the region 
$\Delta_1$
plays an important role when we discuss massless fields.

\subsection{Scalar and spinor fields}

Consider the massless case. Here situation becomes rather
intricate. Integration over interval [0,1] in (\ref{form0}) --- if done
formally --- leads to a result that differs by coefficient 1/2 from what is
to be expected from analogy with the massive case. The difficulty is due to
the fact that all massless states are sitting on the light cone so that
their distribution $\Phi (\rho )\equiv \rho ^2n_iT_{i0}(\rho {\bf n}t,t)$ 
is a 
$\delta -$function. However, the position of the latter corresponds to the
boundary of the integration region at $\rho =1$. Therefore, care is needed
in how one defines the $\delta -$function. The trick we use ensures a
natural connection of the massive and massless cases. Namely, let us extend
the integration region into positive direction over unit interval to 
$1+\varepsilon $ 
and decrease it near zero, obtain: $\rho =\left[
1-\varepsilon ^{\prime },1+\varepsilon \right] ~,\forall \varepsilon
^{\prime }\in (0,1)\ ,\forall \varepsilon \in (0,\infty ).$ 
The non-zero
part of the integral (\ref{form0}) comes from the infinitesimal region near
1 symmetrically on both sides.

All massless particles occupy the infinitesimally narrow $\delta $-region 
at 
$\rho =1.$
The stationary phase method gives the equations:
\begin{equation}
\label{w1}\frac{{\bf p}}{\mid {\bf p}\mid }=
\rho {\bf n,\quad }\frac{{\bf q}}{\mid {\bf q}\mid }=\rho {\bf n}%
, 
\end{equation}

\begin{equation}
\label{w2}\epsilon \cdot \epsilon ^{\prime }{\bf np}+{\bf nq}=0, 
\end{equation}
where $\epsilon $ and $\epsilon ^{\prime }$ independently takes the values 
$\pm 1$. 
For the positive sign of $\epsilon \cdot \epsilon ^{\prime }$, the 
stationary manifold is ${\bf p}={\bf q}%
=0,\left| {\bf p}\right| =p_0=0.$ In the stationary region 
asymptotically 
$\varepsilon \;({\bf n})=0.$

For the negative $\epsilon \cdot \epsilon ^{\prime }$ sign the stationary 
manifold is: $\rho =1,\,{\bf p}=%
{\bf q}=\varpi {\bf n}$, in the stationary region 
\begin{equation}
\label{c15}
{\bf p}^{\prime }=\varpi {\bf n}+\frac{{\bf p}}{\sqrt{t}}%
,\,\;\;{\bf q}^{\prime }=\varpi {\bf n}+\frac{{\bf q}}{\sqrt{t}},\,\quad
\,\rho ^{\prime }=1-\frac \rho {\sqrt{t}}. 
\end{equation}
Here after representing ${\bf p,q}$ as a series 
we use change of variables 
${\bf p\leftrightarrow p}^{\prime },{\bf q\leftrightarrow q}^{\prime 
}$. 
As was announced above, let us introduce $\varepsilon $ into 
integration
limits over $\rho $ $(0<\varepsilon <1).$ 
After substituting the above into 
(%
\ref{form0}), for scalar fields one obtains the following formula: 
$$
\varepsilon \;({\bf n})=\lim \limits_{t\rightarrow \infty }\frac{t^3}{(2\pi
)^3}\int\limits_\varepsilon ^{1+\varepsilon }\rho ^{\prime \;2}d\rho
^{\prime } \ {n_i} \int \frac{d{\bf p}^{\prime }}{2p_0^{\prime }}\int 
\frac{d{\bf 
q}%
^{\prime }}{2q_0^{\prime }}p_0^{\prime }\cdot q_i^{\prime }\cdot\epsilon 
\cdot \epsilon ^{\prime }\cdot
a^\epsilon ({\bf p}^{\prime })a^{\epsilon ^{\prime }}({\bf q}^{\prime }) 
$$
\begin{equation}
\label{c16}\times \exp \left\{ it\;[\epsilon \;(p_0^{\prime }-{\bf p}%
^{\prime }{\bf n}\rho ^{\prime })+\epsilon ^{\prime }\;(q_0^{\prime }-{\bf 
q}%
^{\prime }{\bf n}\rho ^{\prime })]\right\} 
\end{equation}
Let us insert into this expression $1=\int\limits_0^\infty \delta (\varpi
-p)\ d\varpi $. One can represent the absolute value of ${\bf p}$ as series
and retain only contributions of order $O(\frac 1{\sqrt{t}})$. It yields 
the
argument of the $\delta $-function:
$\left( -\frac{(\hbox{\bf p},\hbox{\bf n})}{\sqrt{t}}\right)$. 
 One must take into account 
\begin{equation}
\label{c17}({\bf p}_{\bot },{\bf n})=0,\quad {\bf p}={\bf p}_{\bot }+{\bf 
p}%
_{\Vert }. 
\end{equation}
The integrals over ${\bf p}_{\bot },{\bf q}_{\perp }$ are trivial and 
yield the factor $(2\pi )^2\varpi ^2$. Now one can do the integrals 
over $q_{\Vert},\rho $ 
in two independent ways with the same final result:

The first way is to take the integral over $\rho $ first and apply the 
formula:
\begin{equation}
\label{c19}\int\limits_{-\infty }^\infty dq_{\Vert }\frac{( e^{ibq_{\Vert 
}}- e^{-iaq_{\Vert
}})}{iq_{\Vert }}=2\pi \,. 
\end{equation}
Note that the result is independent of $a,b>0.$

The second way is to take the integral over $q_{\Vert }$ first and then the 
integral over 
$\rho $. 
Take the limit and use the formula: 
\begin{equation}
\label{c20}\int\limits_{-\infty }^\infty d\rho \,\delta (\rho )=1. 
\end{equation}
In both cases one obtains the following result:
\begin{equation}
\label{c21}\frac 14\int\limits_0^\infty \varpi ^2d\varpi
\;(a^{+}a^{-}(\varpi {\bf n)}+a^{-}a^{+}(\varpi {\bf n)}){.} 
\end{equation}

For massless spinors the calculations are similar. 
The result is 
\begin{equation}
\label{c22}\frac 12\int\limits_0^\infty \varpi ^2d\varpi \,\,\left(
b_s^{+}b_s(\varpi {\bf n)}+d_s^{+}d_s(\varpi {\bf n)}\right) \,. 
\end{equation}

\subsection{Gauge field}

The massless vector case is slightly more interesting because it is
necessary to consider the problem due to broken Lorentz
invariance when one deals with physical states. 
It is
impossible to use the local frame of reference in momentum space 
as in the massive case 
because of the 
denominator that is singular at ( $m=0$ ). 
To circumvent the problem one can use the standard Gupta-Bleuler
quantization procedure (see e.g.\ \cite{bog-shir}). 
For quantum fields the Lorentz condition can be rewritten 
as a
condition for physical states: 
\begin{equation}
\label{c2223}\partial _\mu A_\mu ^{-}\Phi =0,\quad \Phi ^{*}\partial _\mu
A_\mu ^{+}=0. 
\end{equation}
Now one turns to the local
frame of reference \cite{bog-shir}. The non-physical zeroth and third 
components cancel each other as expected. 
In the average value over a physical state $\Phi $ only the physical 
components remain. 
(Average value of observable over physical states is the same as over all 
states).
Let us turn to the Lagrangian in Lorentz gauge ($\alpha =0$). 
One takes into
account the above arguments, repeats the 
calculation of $\varepsilon \;({\bf n})$, 
and obtains:
\begin{equation}
\label{c24}\frac 14\int\limits_0^\infty \varpi ^2d\varpi
\;(a_1^{+}a_1^{-}(\varpi {\bf n})+a_2^{+}a_2^{-}(\varpi {\bf n}%
)+a_1^{-}a_1^{+}(\varpi {\bf n})+a_2^{-}a_2^{+}(\varpi {\bf n})). 
\end{equation}

\section{Conclusion}

We have seen that the expression (\ref{form0}) 
of C-correlators in terms of the energy-momenum tensor 
$T_{\mu \nu }$ is correct for non-scalar particles as well. 
The distribution of massive particles lies 
inside the integration region for $\rho $ (see
Fig.1 on the previous page). 

In the massless case the essence of calculation is that all particles are
sitting on the light cone. 
Therefore the operator-valued distribution $\Phi $ is 
$\delta $-function (see Fig.2 above). 
It follows from our analysis that the $\theta $-function 
that describes the integration region must be defined
as a limit ``from the right'' of smooth regulators (Fig.3).

This definition for the massless case should be important:
(i) in axiomatic proofs; 
(ii) in proofs of a generalization of KLN theorem for jet observables
\footnote{This work was planned by N.A.\,Sveshnikov.}.

For gauge particles we verified the formula (\ref{form0}) checked for 
Lorentz gauge using in
the Gupta-Bleuler formalism. 
It would be interesting to develop a treatment for general 
(not necessarily covairant) gauges.
There is systematic formalism for treatment of arbitrary gauges 
for the case of Green's functions (see e.g. \cite{fad-slav}) but not for 
the Fock space. 
It would be useful to develop such a formalism in order to
clarify 
calculational issues enocountered 
when there are loop and phase space integrals
simultaneously.

\section{Acknowledgements}
P.C. thanks F.Tkachov for a help with preparation of this paper 
and 
the organizers of the XII Int. Workshop QFTHEP'97 
for financial support.


\end{document}